\documentclass[preprint,preprintnumbers,nofootinbib]{revtex4}
\usepackage{graphicx}
\usepackage{dcolumn}
\usepackage{bm,amsfonts,amsthm,amsmath,amssymb}
\usepackage[]{epsfig,graphics}
\usepackage{comment}
\usepackage[T1]{fontenc}
\usepackage{lipsum}
\usepackage[utf8]{inputenc}
\usepackage{ulem}
\usepackage{multirow}
\usepackage{color}
\usepackage{lastpage}
\usepackage[sort&compress]{natbib}
\usepackage{enumerate}
\usepackage{subfigure}
\usepackage{wasysym}
\usepackage{hyperref}
\usepackage{caption}
\usepackage{physics}
\usepackage{tensor}
\usepackage{xfrac}

\hypersetup{
    bookmarks=true,         
    unicode=false,          
    pdftoolbar=true,        
    pdfmenubar=true,        
    pdffitwindow=false,     
    pdfstartview={FitH},    
    pdftitle={My title},    
    pdfauthor={Author},     
    pdfsubject={Subject},   
    pdfcreator={Creator},   
    pdfproducer={Producer}, 
    pdfkeywords={keyword1} {key2} {key3}, 
    pdfnewwindow=true,      
    colorlinks=false,       
    linkcolor=red,          
    citecolor=green,        
    filecolor=magenta,      
    urlcolor=cyan           
}

\newenvironment{itemize*}
  {\begin{itemize}
    \setlength{\itemsep}{0pt}
    \setlength{\parskip}{0pt}}
  {\end{itemize}}

\newenvironment{enumerate*}
  {\begin{enumerate}
    \setlength{\itemsep}{0pt}
    \setlength{\parskip}{0pt}}
  {\end{enumerate}}

\newenvironment{description*}
  {\begin{description}
    \setlength{\itemsep}{0pt}
    \setlength{\parskip}{0pt}}
  {\end{description}}

\def\ben{\begin{enumerate*}}
\def\een{\end{enumerate*}}
\def\bi{\begin{itemize*}}
\def\ei{\end{itemize*}}
\def\bd{\begin{description*}}
\def\ed{\end{description*}}
\def\be{\begin{equation}}
\def\ee{\end{equation}}
\def\bea{\begin{eqnarray}}
\def\eea{\end{eqnarray}}
\def\bfl{\begin{flushleft}}
\def\efl{\end{flushleft}}

\begin{document}

\title{Can Cosmology Provide a Test of Quantum Mechanics?}

\author{Julian Georg}
\email{georgj9@rpi.edu}
\affiliation{Department of Physics, Applied Physics, and Astronomy, Rensselaer Polytechnic Institute,\\
110 8th Street, Troy, NY 12180, USA}
\author{Carl Rosenzweig}
\email{crosenzw@syr.edu} 
\affiliation{Department of Physics, Syracuse University, Syracuse, NY 13244, USA}

\date{\today}

\begin{abstract}
Inflation predicts that quantum fluctuations determine the large scale structure of the Universe. This raises the striking possibility that quantum mechanics, developed to describe nature at short distances, can be tested by studying nature at its most immense -- cosmology. We illustrate the potential of such a test by adapting the simplest form of the inflationary paradigm. A nonlinear generalization of quantum mechanics modifies predictions for the cosmological power spectrum. If we assume that the nonlinear parameter $b$ is a comoving quantity observational cosmology, within the context of single field inflation, is sufficiently precise to place a stringent limit, $b\leq 3\times 10^{-37}$ eV, on the current, physical size of the nonlinear term.
\end{abstract}

\maketitle
\thispagestyle{empty}
\section{Introduction}\label{sec:int}
Quantum mechanics may be the most successful theory in all physics. It has been applied successfully to widely diverse situations and each successful prediction in an atomic, or quantum electrodynamic context is of course a test of quantum mechanics. It is, however, difficult to subject the fundamental theory to precision tests.  All theories benefit from having an alternative to serve as a foil. This is neither easy nor commonly done for quantum mechanics.  In the words of Steven Weinberg \cite{WEINBERG1989336},

``Considering the pervasive importance of quantum mechanics in modern physics, it is odd how rarely one hears of efforts to test quantum mechanics experimentally with high precision~.~.~.~it ought to be possible to test quantum mechanics more stringently than any individual quantum theory~.~.~. . Perhaps we can formulate  experiments that would show up departures from quantum mechanics itself.''  
 
It is important to find venues where we can test our most basic assumptions and theories.  The cosmos presents us with one new arena. A spectacular, and naively counter intuitive, prediction of inflation is that the large scale structure of our Universe originates from primordial, microscopic quantum fluctuations of the inflaton field $\phi$ which drives inflation. This is possible because of inflation's ability to stretch small regions of space to enormous size. The picture receives strong support from recent experiments \cite{PLANCKXX}. Data show we are in the age of precision cosmology and confirm, to good accuracy, this cardinal prediction of inflation -- the imprint of quantum mechanics on the Universe.

The successful inflationary, quantum mechanical prediction of the power spectrum offers the novel and surprising possibility that cosmological data can be used to test Quantum Mechanics! To illustrate the feasibility of such a program we accept the single field inflation model and study its predictions for a modification of Quantum Mechanics incorporating a non-linear addition. This modification introduces corrections to the power spectrum.  Cosmological data put a very tight limit on the magnitude of the non-linear term, a limit which exceeds in precision limits derived from table top experiments in the lab. This shows that, in principal, the Universe can provide a precision test of Quantum Mechanics.

We adopt the single field, slow roll model of inflation and start from the action of the scalar inflaton field in a Friedmann Lemaitre Robertson Walker (FLRW) background characterized by the scale factor $a(t)$. We work in flat gauge, follow the standard treatment, and write the inflaton field  as $\phi$ = $\phi_c$ + $\delta\phi(x)$ where $\delta\phi(x)$ is the deviation from a uniform background. The action is then expanded, to second order, in the perturbed inflaton field $v(x)$ with $\delta\phi(x)$ = $\sfrac{v(x)}{a}$.  Slow roll correction terms are neglected. Care must be taken in properly defining variables such as $v_k$ to ensure that we are calculating physical effects rather than gauge artifacts. Finally we perform a Fourier transform to $k$ space and arrive at the classical action
\be\label{eq:action}
S = \frac 12 \int\dd\tau\dd[3]k\left[(v'_k)^2-\left(k^2-\frac{a''}{a} \right)v_k^2\right].
\ee
Primes denote differentiation with respect to conformal time $\tau$ and $\sfrac{\vec{k}}{a}$ is the physical wavenumber (inverse wavelength) of each mode. Details of the steps involved are found in standard treatments of inflationary cosmology.  The equations of motion following from Eq.~(\ref{eq:action}) lead to a simplified Mukhanov-Sasaki \cite{SASAKI1983,KODSAS1984,MUKHANOV1988} equation for $v_k$, \footnote{Because $\phi$ is real, $v_{\vec{k}}^*$ = $v_{-\vec{k}}$ which gives rise to Eq.~(\ref{eq:muksas}) with $v_{\vec{k}}$ dependent only on the magnitude $k$.}
\be\label{eq:muksas}
v_k''+( k^2 -\frac {a''}{a} ) v_k = 0.
\ee
Since $v_k$ is complex Eq.~(\ref{eq:muksas}) represents two equations one each for the real and imaginary parts($v^{\mathfrak{R}}_k$ and $v^{\mathfrak{I}}_k$) of $v_k$. During the early quasi-de Sitter, slow roll phase of inflation, $\sfrac{a''}{a}$ is proportional to $\sfrac{1}{\tau^2}$ becoming very small at early times $\tau\to -\infty$. We recognize Eq.~(\ref{eq:action}) and Eq.~(\ref{eq:muksas}) as those of a classical harmonic oscillator (HO) with time dependent frequency and potential
\be\label{eq:potential}
 V(v_k,\tau,k) = \left(k^2 - \frac{2}{\tau^2} \right )|v_k|^2.
 \ee
 In the limit $\tau\to -\infty$ this becomes a simple harmonic oscillator (SHO). The discussion to this point has been purely classical. 

Eq.~(\ref{eq:muksas}) determines the evolution of the fluctuations but does not set their size. This is where quantum mechanics makes its entrance. In the quantum regime, fluctuations are inevitable and if we have the proper quantum model we can calculate the size of those fluctuations. Think of the SHO where quantum fluctuations of the position around the potential minimum gives  $\expval{x^2}$ = $\frac{\hbar}{2m\omega}$.  Eq.~(\ref{eq:muksas}) embodies the physical picture of quantum fluctuations arising early in the Universe and then growing with inflation only to exit the horizon and freeze out. Eons after inflation has ended these fluctuations re-enter our horizon and begin the process of collapsing into today's structures. 

Quantization is achieved by first defining a canonical momentum and Hamilton with 
\be
p_k =\frac{\delta\mathcal{L}}{\delta v'_k}=v'_k
\ee
giving rise to the Hamiltonian 
\be\label{eq:hamiltonian}
H=\int\dd[3]k\left[p_k^2+v_k^2\left(k^2-\frac{a''}{a}\right)\right]\equiv\int\dd[3]k\mathcal{H}
\ee 
Creation and annihilation operators are then introduced. (Many discussions introduce these operators before the Fourier transform to k space  but this is a matter of choice. We chose to develop the classical picture as far as possible  before introducing quantization.) 

The quantization of the SHO at early times leads to the fluctuations $\langle v_k^2\rangle=\frac {\hbar} {2k}$  fixing the magnitude of $v_k$. The solution to Eq.~(\ref{eq:muksas}) is now  
\be\label{eq:vofk}
v_k(\tau)=\frac{\hbar}{\sqrt[]{2k}}e^{-ik\tau}\left (1-\frac{i}{k\tau}\right).
\ee

An important measure of fluctuations in density is the power spectrum 
\be
\mathcal{P}(k)=k^3\langle |\delta\varphi_k|^2\rangle=\frac{k^3}{a^2}\langle |v_k|^2\rangle|_{\tau\to0}.
\ee
For a de Sitter universe, one finds the scale free behavior 
\be\label{eq:pow}
\mathcal{P}(k)=Ak^{n-1}
\ee
with $n=1$. The coefficient $A$ contains a factor of $\hbar$ making the quantum nature of the prediction explicit. Henceforth we take $\hbar =1$.   Depending on the precise model for inflation there will be corrections to Eq.~(\ref{eq:pow}). Most importantly, slow roll inflation moves $n$ slightly below $1$. The exact shift depends on details of the inflationary potential. The Planck data gives $n=0.9655\pm0.0062$  \cite{PLANCKXX}.

Another property of quantum mechanics needed for the prediction of Eq.~(\ref{eq:pow}) is that fluctuations for different values of $k$ (e.g. $k$ and $k'$) are independent of each other. This property is usually attributed to the linear nature of quantum mechanics.

\section{Nonlinear Quantum Mechanics (NLQM)}
How sensitive is $\mathcal{P}(k)$ to the detailed quantum nature of the fluctuations? One way of answering this question is to use a generalization of quantum mechanics which makes testable predictions for the power spectrum and to compare these predictions to those of standard quantum mechanics.

The logical structure of quantum mechanics is rigid and this rigidity makes it difficult to match its successes with a modified theory.  What can an imagined change or correction to quantum mechanics look like?  Because almost all physical linear theories have, at some level, nonlinear corrections, it is natural to ask if there exists small nonlinear corrections to quantum mechanics. This is more challenging to do than to say, since it is difficult to add nonlinear terms and maintain sensible physical interpretation.  Nevertheless, several authors have tried  \cite{WEINBERG1989336,BIALYNICKIBIRULA1976}.  While there are reasons to be uncomfortable with the nonlinearities (see \cite{GISIN1990,POLCHINSKI1991,JORDAN2009,BASSI2015,HELOU2017} for discussions and many references) physics requires testing not comfort.

The chief source of discomfort is the predicted existence of superluminal signaling \cite{GISIN1990,POLCHINSKI1991} although there are claims (see  \cite{JORDAN2009} for discussion and further references) that this is not immediately disqualifying. Furthermore unpalatable consequences should be subject to experimental tests and we know of no high precision tests ruling out superluminal signalling arising from nonlinearities. For example, Gisin's experiment (see  discussion in  \cite{JORDAN2009}) showing superluminal signaling would take an extremely long time to conduct given present limits on the non linear parameters. It is therefore interesting to examine what cosmology has to say about this. Remarkably the simplest picture of inflation says something significant.

Bialynicki-Birula and Mycielski were able to formulate a nonlinear generalization of the Schr\"odinger equation with an acceptable interpretation \cite{BIALYNICKIBIRULA1976} . They suggested replacing the standard Schr\"odinger equation with the following nonlinear version.  
\begin{equation}
\frac{\hbar} {2m}\nabla^2\Psi(\textbf{r},t) -V(x,t) \Psi -b\Psi(\textbf{r},t) \mathrm{ln} (\vert \Psi (\textbf{r},t)\vert ^2d) = -i\hbar\partial_t\Psi(\textbf{r},t)\label{eq:se}
\end{equation}
The constant $b$, with dimensions of energy, is a universal, positive constant, while $d$, with dimensions such that $\Psi^2d$ is dimensionless, has no physical significance and only adds a phase to the wavefunction. This choice of nonlinearity preserves several desirable properties of quantum mechanics, i.e. factorization of wavefunctions, existence of a lower energy bound and the Planck relation $E=\hbar\omega$. There are also Gaussian solutions.  A general pathology of nonlinear adjustments to the Schr\"odinger equation is that noninteracting particles will influence each other \cite{HOHSU2015}. The logarithmic addition is unique in allowing factorization of the nonlinear Schr\"odinger equation for two noninteracting, nonentangled particles, thus avoiding this pathology. This independence manifests itself in the condition $\expval{v_kv_{k'}}$ proportional to $\delta_{kk'}$. (For the remainder of this discussion we adopt quantization in a box rather than in the continuum.)

Experiments were performed \cite{SHIMONY1979,SHULL1980,KLEIN1981} to find limits on the parameter $b$ with the most stringent limits establishing $b$ < $3\times10^{-15}$ eV \cite{KLEIN1981}.

The technical task confronting us is to calculate $\langle |v_k|^2\rangle$ in this nonlinear generalization of quantum mechanics. Because this generalization relies on the Schr\"odinger equation rather than creation and annihilation operators (as is most common in cosmology literature), we quantize in the Schr\"odinger Picture. The Schr\"odinger approach has been used in cosmology by J. Martin \cite{MARTIN2007} where many calculational details and an extensive set of references can be found. See also \cite{KIEFER1992} for earlier relevant work. The Schr\"odinger picture was used in an interesting attempt to find cosmological limits for theories of wave function collapse by comparing collapse predictions to the power spectrum. See \cite{MARTIN2012}, which includes further references on this topic.

\section{Inflaton Fluctuations in Nonlinear Quantum Mechanics}
Our strategy is to quantize the fluctuations $v_k$ by formulating a Schr\"odinger equation for $\Psi_k(v_k,\tau)$ following \cite{MARTIN2007}. We then modify this Schr\"odinger equation by adding the non-linear term  $-b\Psi(v_k,\tau) \mathrm{ln} \vert \Psi (v_k,\tau)\vert ^2 $.  

We start with the standard action Eq.~(\ref{eq:action}) but now perform the Fourier decomposition in a box.
\be\label{eq:action2}
S = \frac 12 \int\dd\tau\sum_k\left[(v'_k)^2-\left(k^2-\frac{a''}{a} \right)v_k^2\right].
\ee We begin quantization by treating $v_k$ and $p_k$ as operators and imposing the standard commutation relation between $\hat p_k$ and $\hat v_k$
\be
\left [ \hat v_k,\hat p_{k'}\right ] =i\delta_{kk'},
\ee
which leads to the standard representation
\be
\hat v_k\Psi_k=v_k\Psi_k, \quad \hat p_k\Psi_k=-i\frac{\partial\Psi_k}{\partial v_k}.
\ee
The Schr\"odinger equation then follows 
\begin{equation}\label{eq:schroed}
i\partial_{\tau}\Psi_k(v_k,\tau)=\hat{\mathcal{H}}\Psi_k(v_k,\tau)=\left [ (\hat p_k)^2 + V(v_k,\tau,k)\right ] \Psi_k(v_k,\tau),
\end{equation}
We exploited the property of the independence of noninteracting oscillators.  This property is rare in arbitrary nonlinear additions to the Schr\"odinger equation, but is satisfied by the nonlinear term of Birula et al. Not surprisingly, given the formal similarity between the perturbative inflaton action and the action for the HO, Eq.~(\ref{eq:schroed}) is essentially identical to the non-relativistic Schr\"odinger equation for the HO and for which we have a recipe for modification, Eq.~(\ref{eq:se}).  According to \cite{BIALYNICKIBIRULA1976} the intervention for nonlinearization occurs at this stage by adding the suggested nonlinear term, $b\Psi_k(v_k,\tau) \mathrm{ln} \vert \Psi_k(v_k,\tau) \vert ^2$, to the standard Schr\"odinger equation. The final Schr\"odinger equation, including the nonlinear term, is
\begin{equation}
\nabla_{v_k}^2\Psi_k(v_k,\tau) -V(v_k,\tau) \Psi_k(v_k,\tau) -b\Psi_k(v_k,\tau) \mathrm{ln} \vert \Psi_k(v_k,\tau) \vert ^2  d = -i\hbar\partial_{\tau}\Psi(v_k,\tau)\label{eq:fullSE}
\end{equation}

We need to calculate
\be
\langle |v_k(\tau)|^2\rangle_0=\int\mathrm{d}v_k\Psi_k^* \left [ \left ( v^{\mathfrak{R}}_k\right )^2+\left(v^{\mathfrak{I}}_k\right)^2\right ]\Psi_k.
\ee
Since the solution to Eq.~(\ref{eq:fullSE}) with $b=0$ is a Gaussian and the solution for a nonzero $b$ but $V(v_k,\tau)$  independent of $\tau$, is also a Gaussian it is natural, especially for small $b$, to try a Gaussian as a solution for the full Eq.~(\ref{eq:fullSE}). We thus make the Gaussian ansatz for $v^{\mathfrak{R}}_k$ and $v^{\mathfrak{I}}_k$ separately and omit the superscripts from now on.
\be
\Psi_k(\tau, v_k)= N(\tau)\exp(-\Omega(\tau)(v_k)^2).
\ee
with $\Omega(\tau) =g(\tau) + i h(\tau)$. Normalization of $\Psi$ requires $N=\left(\frac{2g}{\pi}\right)^{\sfrac14}$, consistent  \cite{MARTIN2007,KIEFER1992} with the Schr\"odinger equation, giving 
\be
\langle |v_k|^2\rangle_0=\frac{1}{4g}.
\ee
Eq.~(\ref{eq:fullSE}) tells us that the functions $g$ and $h$ satisfy
\begin{subequations}\label{eq:de}
\begin{eqnarray}
h^{\prime}&=&2(h^2-g^2)+\frac 12 k^2-\frac{1}{\tau^2}+2bg, \\
g^{\prime}&=&4gh. \label{eq:de_g}
\end{eqnarray}
\end{subequations}

These equations do not, as far as we could determine, allow a closed solution but we can check interesting limits. If $b =0$ the solutions   $g_0=\frac{k^3\tau^2}{2(1+k^2\tau^2)}$ and $h_0=\frac{1}{2\tau(1+k^2\tau^2)}$ reproduce the well known results for $\langle v_k^2\rangle$ obtained from Eq.~(\ref{eq:vofk}). For early times $\tau\rightarrow -\infty$ the solution to Eqs.~(\ref{eq:de}) is $h = 0$ and $g =\frac12\left(b+\sqrt{b^2+k^2}\right)$.  The choice of this solution to the quadratic equation for $g$ is dictated by imposing the boundary condition that $b\to0$ is the standard SHO result. This leads to 
\be
\Psi(v,\tau)=e^{(-iE\tau)}\left(\frac{g}{\pi}\right)^{\frac{1}{4}}e^{\left(-v^2\left(\sqrt[]{b^2 +k^2}+b\right)\right)},
\ee
where 
\be
E=\frac{b+\sqrt[]{b^2+k^2}}{2} -\frac{b}{2}\left(\ln\pi\left(b+\sqrt[]{b^2+k^2}\right)\right) 
\ee 
is the ground state energy of the SHO in the nonlinear theory \cite{BIALYNICKIBIRULA1976}. Thus we start out in the appropriate vacuum (lowest energy) state .

The power spectrum is determined by $\tau\rightarrow0$ and, since $b$ is very small, we approximate the solutions to Eqs.~(\ref{eq:de}) as a power series in b, $g =g_0 +bg_1 +...$ and $h = h_0 + b h_1 +..$ To first order in $b$ we find  

\be
g(\tau)\xrightarrow[b\rightarrow 0]{} \frac 12 \frac{k^3\tau^2}{k^2\tau^2+1}\left(1+\frac bk+4\frac bk\int\limits_{-\infty}^{\tau}h_1(t)\dd{t}\right),
\ee
which is readily seen to satisfy Eq.~(\ref{eq:de_g}).

Numerical integration of the system of Eqs.~(\ref{eq:de}) gives
\be
g(\tau)\xrightarrow[\tau\rightarrow 0]{} \frac{k^3\tau^2}{2}\left(1+1.7\left(\frac bk\right)\right)
\ee
leading to the prediction for the power spectrum 
\be\label{eq:powmod}
\mathcal{P}(k)=Ak^{n-1}\frac{1}{1+1.7\frac{b}{k}}.
\ee
Slow roll corrections will be the same as for the standard treatment.

\begin{figure}
\begin{center}
\includegraphics[scale=0.8]{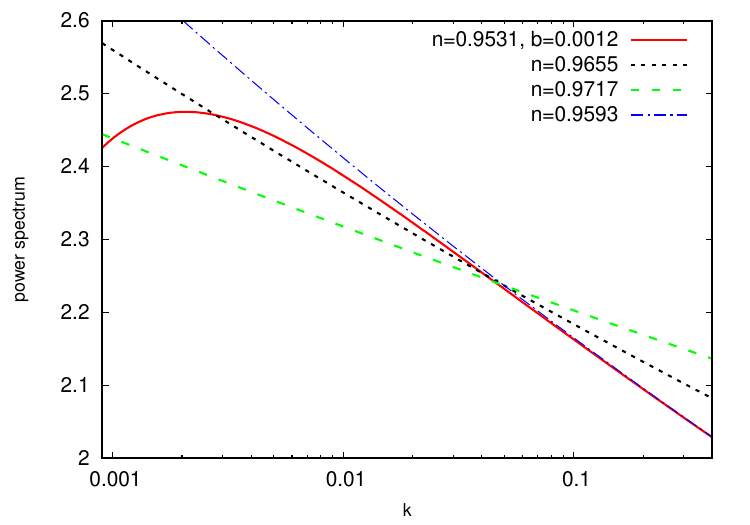}
\end{center}
\caption{\label{fig:1} 
The modified power spectrum, Eq.~(\ref{eq:powmod}), is red (solid) and the $2\sigma$ errors to the Planck value of $n$ Eq.~(\ref{eq:pow}), are blue (dash dot) and green (dash). The central value for n is black (dotted). For $b =0.001$ the modified spectrum just fits inside the $2\sigma$ limits}
\end{figure}

We estimate the constraint on $b$ by first rewriting the power spectrum as 
\be
\mathcal{P}(k)=A\left(\frac{k}{k_*}\right)^{n-1}\frac{1}{1+1.7\left(\frac{b_{\mathrm{cmb}}}{k_*}\right)\left(\frac{k_*}{k_{\mathrm{cmb}}}\right)}.
\ee
The parameter $k_*$ is the (physical) pivot scale for fits to $\mathcal{P}(k)$ and the (physical) quantities $b_{\mathrm{cmb}}$ and $k_{\mathrm{cmb}}$ correspond to $\sfrac{b}{a_{\mathrm{cmb}}}$ and $\sfrac{k}{a_{\mathrm{cmb}}}$ respectively.  $a_{\mathrm{cmb}}$ is the FLRW scale factor at recombination. $k_{\mathrm{cmb}}$ is the physical wavenumber as measured at recombination We then compare. Eq.~(\ref{eq:powmod}) with the Planck data   via Eq.~(\ref{eq:pow}). Our limit on $b_{\mathrm{cmb}}$ is the largest value of $b_{\mathrm{cmb}}$ for which Eq.~(\ref{eq:powmod}) sits within the $2\sigma$ variation on $n$, $n=0.9655\pm0.0124$.  We used the same pivot scale $k_* = 0.05$ Mpc$^{-1}$ and normalized the modified power spectrum to agree with the Planck parameterization at $k_*$. This results in $\sfrac{b_{\mathrm{cmb}}}{k_*} \leq 1\times 10^{-3}$  see Fig.~(\ref{fig:1}), setting a limit

\be\label{eq:bcmb}
b_{\mathrm{cmb}}\leq3\times 10^{-34} \mathrm{\ eV},
\ee
compared to the best terrestrial limit of $3\times10^{-15}$ eV.

What is the interpretation of  Eq.~(\ref{eq:bcmb})? (recall $b_{\mathrm{cmb}} = \sfrac{b}{a_{\mathrm{cmb}}}$)  Eq.~(\ref{eq:powmod}) is unusual for a measurable cosmological quantity.  $k$, appearing in the denominator, is the co-moving wavenumber not the physical wavenumber. Thus we have to confront the nature of the parameter $b$ (with dimension of energy). We consider two possibilities ($i$) $b$ appearing in Eq.~(\ref{eq:fullSE}) is a new, universal, fundamental constant of nature which introduces a new scale into physics and ($ii$) $b$ is fundamental but is a comoving quantity. We treat these in turn.

In case ($i$) the power spectrum, Eq.~(\ref{eq:powmod}), is a function of the ratio of a physical quantity with a co-moving quantity.  This is usually considered inadmissible but in our case it has a physical interpretation.  Because $b$ is a fundamental physical constant which breaks diffeomorphism invariance (see below) it introduces a new scale into physics.  This allows for extra time and scale dependence  to enter. (A similar result is obtained in \cite{MARTIN2012}  with respect to the new fundamental constant which governs wave function collapse.) The power spectrum is now time dependent (through its dependence on $a(t)$) and in principle allows for the determination of the absolute scale factor of expansion in units of our new constant $b$. If we have an independent measurement of $b$, say from a table top quantum mechanical experiment, Eq.~(\ref{eq:bcmb}) provides a lower bound on $a_{\mathrm{cmb}}$ in units of the fundamental length $\sfrac{1}{b}$. If the power spectrum fixes  a non zero value for $\sfrac{b}{a_{\mathrm{cmb}}}$ we have a determination of  $a_{\mathrm{cmb}}$  in units $\sfrac{1}{b}$ If the only data point we have is  Eq.~(\ref{eq:bcmb}) we cannot put a limit on $b$ but only on $\sfrac{b}{a_{\mathrm{cmb}}}$.

Now consider possibility ($ii$) that $b$ is comoving.  The formal reason for allowing only the physical wavenumber to enter into observables is diffeomorphism invariance, an important requirement of General Relativity. The appearance of factors of $\sfrac{b}{k}$ in Eq.~(\ref{eq:powmod}) seem to violate diffeomorphism invariance which, in its simplest form, asserts that physics should be invariant under the rescaling $a\to\lambda a$ and $x\to\lambda^{-1}x$. We are considering changes to quantum mechanics, but are wary about also introducing changes to General Relativity. Therefore, it is worthwhile to see if we can maintain diffeomorphism invariance.

To analyze this, we examine the properties of $b$ under such transformations. We insist that the action for each $k$ mode $S_k=\int\dd{\tau}\dd{v_k}\mathcal{L}_k$, from which the modified Schr\"odinger Eq.~(\ref{eq:fullSE}) can be derived, is diffeomorphism invariant. $\mathcal{L}_k$, the Lagrangian density, is given by
\begin{equation}
\mathcal{L}_k=\Psi_k^*\frac{\partial\Psi_k}{\partial\tau}-\frac{\partial\Psi_k^*}{\partial v_k}\frac{\partial\Psi_k}{\partial v_k}+V(v_k,\tau,k)|\Psi_k|^2 -b|\Psi_k|^2\mathrm{ln}(|\Psi_k|^2d).
   \end{equation} 
$\tau$, $k$, and $\phi(x)$, the inflaton field, are assumed to have the standard variations under $a\to\lambda a$ i.e. $k\to\lambda k$,  $\tau\to\lambda^{-1}\tau$ and $\phi(x)\to\phi(\lambda^{-1}x)$. The properties of $v_k$ and $\Psi_k$ can be determined by the definition of $v_k$ in terms of $\delta\phi_k$ and the normalization of $\Psi_k$.  This implies that $v_k\rightarrow\lambda^{-\sfrac 12}v_k$ and $\Psi_k\to \lambda^{\sfrac14} \Psi_k$. The action will be invariant if $\mathcal{L}_k\to\lambda^{\sfrac32} \mathcal{L}_k$ which will be true if $b\to\lambda b$. 

This means that $b$ is not a true constant but is itself a comoving quantity. This is unusual but conceivable. For instance $b$ could be proportional to $\sqrt{K}$ where $K$ is the Gaussian curvature of the Universe.  It would thus be comoving and naturally exceedingly small at the present time. One could imagine that the nonlinear term containing $b$ somehow arises from quantization of gravity. (The so called constant $b$ varying  with the size of the Universe is also reminiscent of Dirac's large number hypothesis).

How does this limit relate to the laboratory tests of NLQM?  In order to maintain diffeomorphism invariance  we chose to treat the cosmological application of NLQM as the fundamental appearance of the NLQM parameter $b$ on the physical stage. We modified Eq.~(\ref{eq:schroed}), the appropriate Schr\"odinger equation for the quantization of the inflaton fluctuations, by adding the nonlinear term 
$-b\Psi_k(v_k,\tau) \mathrm{ln} \vert \Psi_k(v_k,\tau) \vert ^2d$.  Maintaining diffeomorphism invariance required $b$ to be a comoving quantity. From this point of view the non-relativistic Eq.~(\ref{eq:se}) is now an approximation with $b$, along with all other quantities appearing in Eq.~(\ref{eq:se}), physical.  Thus it will appear in Eq.~(\ref{eq:se}) as $b_{now}$ or $\sfrac{b}{a_{\mathrm{now}}}$. For terrestrial time scales, $a$ is essentially constant and so $b_{\mathrm{now}}$  is a constant as required by the interpretation of Eq.~(\ref{eq:se}) as a candidate NLQM. Since ${a_{\mathrm{now}}}$ is about 1000 times $a_{\mathrm{cmb}}$, $b_{\mathrm{now}}$ is $\sfrac{1}{1000}$
$b_{\mathrm{cmb}}$. From Eq.~(\ref{eq:bcmb}).

\be\label{eq:bnow}
b_{\mathrm{now}}\leq3\times 10^{-37} \mathrm{\ eV},
\ee

\section{Summary and Conclusion}
The purpose of this note is not to challenge quantum mechanics but to celebrate modern cosmology.  The strides made in experimental and theoretical cosmology in the past decades have made it a precision science.  This precision strongly re-enforces the inflationary paradigm whose most spectacular result is that the large scale structure of the Universe is determined by small scale quantum fluctuations.  To highlight this striking prediction we made the optimistic assumption that simple inflation is strictly true and used it to \textit{test} quantum mechanics. If we further assume that the fundamental nonlinear parameter $b$ is comoving we obtain an extremely tight restriction on a specific nonlinear generalization of quantum mechanics that far exceeds precision laboratory experiments. We find it highly noteworthy that measurements of large sale structure of the Universe, within a well defined, if idealized, theoretical framework, provide a precision test of quantum mechanics. The connection between the Universe and the quantum world is quite tight.\\ 

\section*{Acknowledgements}
We would like to thank S.~Watson, R.~Penco, A.~Beiter, A.~Srivastava and our colleagues at Syracuse for discussions. Special thanks are due to C.~Armendariz-Picon for enlightenment and encouragement.

\bibliographystyle{plain}

\end{document}